\documentclass[10pt,twoside]{epnt1p}
\usepackage{epsfig}

\usepackage[normalem]{ulem}
\usepackage{graphicx}
\usepackage{array}
\usepackage{amssymb}
\usepackage{amsmath}

\usepackage{url}

\newcommand{\Dmq}{\Delta m^2}
\newcommand{\Dlt}{\Delta\delta}

\newcolumntype{C}{>{~$}c<{$~}}
\newcolumntype{R}{>{~$}r<{$~}}

\begin{document}

%%%%%%%%%%%%%%%%%%%%%%%%%%%%%%%%%%%%%%%%%%%%%%%%%%% Title, authors and addresses
\begin{frontmatter}


\title{Non-standard Neutrino Oscillations at Icecube}

% use optional labels between square brackets to link authors explicitly to addresses:
% If more than one author, keep a comma between the author tags

\author[address1]{M.C. Gonzalez-Garcia},
\address[address1]{
  Instituci\'o Catalana de Recerca i Estudis Avan\c{c}ats (ICREA), \\
  Departament d'Estructura i Constituents de la Mat\`eria,
  Universitat de Barcelona,\\
  Diagonal 647, E-08028 Barcelona, Spain\\
{\rm and:}  C.N. Yang Institute for Theoretical Physics\\
  Stony Brook University, 
  Stony Brook, NY 11794-3840, USA}
\begin{abstract}
In this talk I review the potential of Icecube for revealing 
physics beyond the standard model in the oscillation of
atmospheric neutrinos~\cite{ouricecube}.
\end{abstract}

% \begin{keyword}

%\PACS 
% \end{keyword}

\end{frontmatter}

\section{Introduction}
With its high statistics data~\cite{skatmlast}
Super--Kamiokande (SK) established beyond doubt that the observed
deficit in the $\mu$-like atmospheric events is due to oscillations, a
result supported by the K2K and MINOS long-baseline (LBL)
experiments~\cite{k2kprl,MINOSdat}.

Mass oscillations are not the only possible
mechanism for atmospheric (ATM) $\nu_\mu \to \nu_\tau$ flavour
transitions. These can be also generated by a variety
of nonstandard physics characterized by the presence of
an unconventional $\nu$ interaction 
that mixes neutrino flavours~\cite{npreview}.  Examples include
violations of the equivalence principle (VEP)~\cite{VEP,qVEP},
non-standard neutrino interactions with matter~\cite{NSI}, neutrino
couplings to space-time torsion fields~\cite{torsion}, violations of
Lorentz invariance (VLI)~\cite{VLI1} and of CPT
symmetry~\cite{VLICPT1,VLICPT2}. 
In contrast to the $E$ energy dependence of the conventional
oscillation length, new physics can produce neutrino oscillations with
wavelengths that are constant or decrease with energy. 
~\cite{yasuda1,flanagan}. At present these scenarios cannot explain the
data\cite{oldatmfitnp} and a combined
analysis of the ATM and LBL data can be performed to 
constraint them even as  subdominant
oscillation effects~\cite{ouratmnp}.

IceCube, with energy
reach in the $0.1 \sim 10^4$\,TeV range for ATM neutrinos, is
the ideal experiment to search for new physics. For most of this
energy interval standard $\Delta m^2$ oscillations are suppressed and
therefore the observation of an angular distortion of the ATM
neutrino flux or its energy dependence provide a clear signature for
the presence of new physics mixing neutrino flavours~\cite{ouricecube}.

\section{Propagation in Matter of High Energy  Oscillating Neutrinos}
\label{sec:formaprop}
We concentrate on $\nu_\mu$--$\nu_\tau$ flavour mixing
mechanisms for which the propagation of 
$\nu$'s ($+$) and
$\bar\nu$'s ($-$) is governed by the following
Hamiltonian~\cite{VLICPT2}:
\begin{eqnarray} \label{eq:hamil}
    {\rm H}_\pm \equiv
    \dfrac{\Dmq}{4 E}
    \mathbf{U}_\theta
    \begin{pmatrix}
	-1 & ~0 \\
	\hphantom{-}0 & ~1
    \end{pmatrix}
    \mathbf{U}_\theta^\dagger
    + \sum_n
    \sigma_n^\pm \dfrac{\Dlt_n\, E^n}{2}
    \mathbf{U}_{\xi_n,\pm\eta_n}
    \begin{pmatrix}
	-1 & ~0 \\
	\hphantom{-}0 & ~1
    \end{pmatrix}
    \mathbf{U}_{\xi_n,\pm\eta_n}^\dagger \;, 
\end{eqnarray}
$\Dmq$ is the mass--squared difference between the two neutrino
mass eigenstates, $\sigma_n^\pm$ accounts for a possible relative
sign of the new physics (NP) 
effects between $\nu$'s and $\bar\nu$'s and
$\Dlt_n$ parametrizes the size of the NP terms. 
By $\eta_n$ we denote the possible non-vanishing relative phases. 
\begin{eqnarray}
\label{eq:rotat}
    \mathbf{U}_\theta =
    \begin{pmatrix}
	\hphantom{-}\cos\theta & ~\sin\theta \\
	-\sin\theta & ~\cos\theta
    \end{pmatrix}\,,
    \qquad
    \mathbf{U}_{\xi_n,\pm\eta_n} =
    \begin{pmatrix}
	\hphantom{-}\cos\xi_n\hphantom{e^{-i\eta_n}} 
	& ~\sin\xi_n e^{\pm i\eta_n} 
	\\
	-\sin\xi_n e^{\mp i\eta_n} 
	& ~\cos\xi_n\hphantom{e^{-i\eta_n}}
    \end{pmatrix}\,;
\end{eqnarray}

If NP strength is constant along the neutrino trajectory the oscillation
probabilities take the form ~\cite{VLICPT2}:
\begin{eqnarray}
&& \!\!\!\!\!\! \!\!\!\!\!\!  \!\!  
P_{\nu_\mu \to \nu_\mu} = 1 - P_{\nu_\mu \to \nu_\tau} =
    1 - \sin^2 2\Theta \, \sin^2 \left( 
    \frac{\Dmq L}{4E} \, \mathcal{R} \right)
\nonumber    \\
&&    
\!\!\!\!\!\!\!\!\!\!\!\!\!\!      
\sin^2 2\Theta = \frac{1}{\mathcal{R}^2} \left(
    \sin^2 2\theta + R_n^2 \sin^2 2\xi_n
    + 2 R_n \sin 2\theta \sin 2\xi_n c\eta_n \right) \,,
\nonumber    \\
&&    
\!\!\!\!\!\!\!\!\!\!\!\!\!\!      
\mathcal{R} =\sqrt{1 + R_n^2 + 2 R_n \left( \cos 2\theta \cos 2\xi_n
      + \sin 2\theta \sin 2\xi_n c\eta_n \right)}\;, 
\;\;\;\;
    R_n = \sigma_n^+ \frac{\Dlt_n E^n}{2} \, \frac{4E}{\Dmq} \,,
\nonumber 
\end{eqnarray}
where, for simplicity, we have assumed  scenarios 
with one NP source characterized by a unique $\Delta\delta_n$.
$c\eta_n=\cos\eta_n$ 

Eq.~\eqref{eq:hamil} describes, for example, flavour mixing due to
new tensor-like interactions for which $n=1$ leading to a
contribution to the oscillation wavelength inversely proportional to 
the neutrino energy. This is the case for 
$\nu_\mu$'s and $\nu_\tau$'s of different masses in the
presence of violation of the equivalence principle  due to non-
universal coupling of the neutrinos, $\gamma_1\neq \gamma_2$
to the gravitational potential
$\phi$~\cite{VEP},  
so  $\Dlt_1 = 2 |\phi|(\gamma_1- \gamma_2) \equiv 2 |\phi| \Delta\gamma $.
$\nu_1$ and $\nu_2$ are related to $\nu_\mu$ and $\nu_\tau$ by a rotation
$\xi_1=\xi_{ vep}$.

For constant potential $\phi$, this mechanism is phenomenologically
equivalent to the breakdown of Lorentz invariance resulting from different
asymptotic values of the velocity of the neutrinos, $c_1\neq c_2$,
$ \Dlt_1 = (c_1- c_2)\equiv\delta c/c$, 
with $\nu_1$ and $\nu_2$ being related to $\nu_\mu$ and $\nu_\tau$ by
a rotation $\xi_1=\xi_{vli}$~\cite{VLI1}.  

For vector-like interactions, $n=0$,  the oscillation wavelength 
is energy-independent. This
may arise, for instance, from a non-universal coupling of the
neutrinos, $k_1\neq k_2$ so $\Dlt_0= Q (k_1- k_2)$
($\nu_1$ and $\nu_2$ is related to the
$\nu_\mu$ and $\nu_\tau$ by a rotation $\xi_0=\xi_Q$), to a space-time
torsion field $Q$~\cite{torsion}.
Violation of CPT resulting from Lorentz-violating effects such 
as the operator,  
$\bar{\nu}_L^\alpha b_\mu^{\alpha\beta} \gamma_\mu
\nu_L^\beta$,  also leads to an
energy independent contribution to the oscillation
wavelength~\cite{VLICPT1,VLICPT2}  which is a function of 
the eigenvalues of the Lorentz violating CPT-odd
operator, $b_i$, $
\Dlt_0 = b_1-b_2 $,
and  the rotation angle, $\xi_0=\xi_{\not\text{CPT}} $, between the
corresponding eigenstates $\nu_i$ and the flavour states
$\nu_\alpha$.

%At present the strongest limits on NP neutrino oscillations
%arise from the non-observation of departure from the $\Delta m^2$
%oscillation behaviour in ATM neutrinos at SK and the
%confirmation of $\nu_\mu$ oscillations with the same 
%oscillation parameters from LBL experiments~\cite{ouratmnp}.

For most of the neutrino energies considered here, 
$\Delta m^2$ oscillations are suppressed and the NP effect is
directly observed. Thus the results will be independent
of the phase $\eta_n$ and we can chose the NP parameters in the range
$\Dlt_n \geq 0$,   $0 \leq \xi_n \leq \pi/4 $.

The Hamiltonian of Eq.~(\ref{eq:hamil}) describes the coherent
evolution of the $\nu_\mu$--$\nu_\tau$ ensemble for any neutrino
energy.  High-energy neutrinos propagating in the Earth can also interact
inelastically with the Earth matter either by charged current (CC)
and neutral current (NC) and as a
consequence the neutrino flux is attenuated.  This attenuation is
qualitatively and quantitatively different for $\nu_\tau$'s and
$\nu_\mu$'s. $\nu_\mu$'s are absorbed by CC  interactions while 
$\nu_\tau$'s are regenerated because they produce a $\tau$ that
decays into another tau neutrino
before losing energy ~\cite{hs}. As a consequence, for each $\nu_\tau$
lost in CC interactions, another $\nu_\tau$ appears (degraded in
energy) from the $\tau$ decay and the Earth never becomes opaque to
$\nu_\tau's$.  Furthermore, a 
secondary flux of $\bar\nu_\mu$'s is also generated in the leptonic
decay $\tau\rightarrow \mu\bar\nu_\mu\nu_\tau$~\cite{kolb}.

Attenuation and regeneration effects of incoherent neutrino fluxes can
be consistently described by a set of coupled partial
integro-differential cascade equations (see for example~\cite{reno}
and references therein) or by a Monte Carlo simulation of the neutrino
propagation in matter~\cite{hs,kolb,crotty}. For astrophysical $\nu$'s, 
because of the long distance traveled  
from the source, the oscillations average out and at the Earth the neutrinos 
can be treated as an incoherent superposition of
mass eigenstates.

For ATM neutrinos this is not the case because oscillation,
attenuation, and regeneration effects occur simultaneously when the
neutrino beam travels across the Earth's matter. 
For conventional neutrino oscillations this
fact can be ignored because the neutrino energies covered by current
experiments are low enough for attenuation and regeneration effects to
be negligible. But for non-standard scenario oscillations,
future experiments probe high-energy neutrinos for which the
attenuation and regeneration effects have to be accounted for
simultaneously. 

In order to do so it is convenient to use the density matrix 
formalism to describe neutrino flavour oscillations. 
The evolution of the neutrino ensemble is determined
by the Liouville equation for the density matrix 
$\rho(t)=\nu(t)\otimes \nu(t)^\dagger$
\begin{equation}
\frac{d{\rho}}{dt}=-i[{\rm H}, {\rho}] \, ,
\end{equation}
where ${\rm H}$ is given by Eq.~(\ref{eq:hamil}). The survival probability
is  given by 
$P_{\mu\mu}(t)={\rm Tr}[\Pi_{\nu_\mu}\,\rho(t)]$, where  
$\Pi_{\nu_\mu}=\nu_\mu\otimes \nu_\mu$ is the $\nu_\mu$ state projector, 
and with initial condition  $\rho(0)=\Pi_{\nu_\mu}$. An equivalent 
equation can be written for the $\bar\nu$ density matrix.

In this formalism attenuation effects due to CC and 
NC interactions 
can be introduced by relaxing the condition ${\rm Tr}(\rho)=1$. 
In this case 
\begin{equation}
\frac{d{\rho(E,t)}}{dt}=-i[{\rm H}(E), \rho(E,t)]
-\sum_\alpha \frac{1}{2\lambda^\alpha_{\rm int}(E,t)}
\left\{\Pi_\alpha,\rho(E,t)\right\} \, ,
\end{equation}
where $[\lambda^\alpha_{int}(E,t)]^{-1}\equiv
[\lambda^\alpha_{\rm CC}(E,t)]^{-1}+
[\lambda_{\rm NC}(E,t)]^{-1}$, $[\lambda^\alpha_{\rm CC}(E,t)]^{-1}
=n_T(x)\, \sigma^{\alpha}_{\rm CC}(E)$, and 
$[\lambda_{\rm NC}(E,t)]^{-1}=n_T(x)\, \sigma_{\rm NC}(E)$
($n_T(x)$ is the number density of nucleons at the point $x=ct$). 

$\nu_\tau$ regeneration and neutrino energy degradation can be accounted
for by coupling these equations to the  shower equations for the $\tau$ flux,
$F_\tau(E_\tau,t)$ (we denote by $F$ the differential fluxes 
$d\phi/(dE \, d\cos\theta)$).
For convenience we define the {\sl neutrino flux density matrix} 
$F_\nu(E,x)=F_{\nu_\mu}(E,x_0)\rho(E,x=c\,t)$   
where $F_{\nu_\mu}(E,x_0)$ is the initial neutrino flux:
\begin{eqnarray}
\frac{d{F_\nu}(E_\nu,x)}{dx}
&=&-i[{\rm H}, F_\nu(E_\nu,x)]
-\sum_\alpha \frac{1}{2\lambda^\alpha_{\rm int}(E_\nu,x)}
\left\{\Pi_\alpha,F_\nu(E_\nu,x)\right\}\nonumber \\
& &+ 
\int_{E_\nu}^\infty  \frac{1}{\lambda_{\rm NC}(E'_\nu,x)}
F_\nu(E'_\nu,x) 
\frac{d N_{\rm NC}(E'_\nu,E_\nu)}{d E_\nu} dE'_\nu  \nonumber \\
& & +
\int_{E_\nu}^\infty   \frac{1}{\lambda^\tau_{\rm dec}(E_\tau,x)}
 F_\tau(E_\tau,x) 
 \frac{d N_{\rm dec} (E_\tau,E_\nu)}{d E_\nu} dE_\tau\, \Pi_\tau \nonumber \\
& & + 
{\rm Br_{\mu}}\,
\int_{E_\nu}^\infty   
\frac{1}{\lambda^\tau_{\rm dec}(E_\tau,x)}\, 
\bar{F}_\tau(\bar E_\tau,x) 
 \frac{d \bar{N}_{\rm dec} 
(\bar{E}_\tau,E_\nu)}{d E_\nu} d\bar{E}_\tau\, \Pi_\tau \, , 
\label{eq:nshower}  \\
\frac{d F_\tau(E_\tau,t)}{d\,x}&=&-\frac{1}{\lambda^\tau_{dec}(E_\tau,x)}
F_\tau(E_\tau,x) \nonumber \\
&&
+ 
\int_{E_\tau}^\infty 
\frac{1}{\lambda^\tau_{\rm CC}(E_\nu,t)} {\rm Tr}[\Pi_\tau\, {F_\nu}(E_\nu,t)] 
 \frac{d N_{\rm CC}(E_\nu,E_\tau)}{d E_\tau} d E_\nu\, .  
\label{eq:tshower}
\end{eqnarray}
$\lambda^\tau_{\rm dec}(E_\tau,x)=\gamma_\tau\, c\, \tau_\tau$. 
$\tau_\tau$ is the $\tau$ lifetime and  
$\gamma_\tau=E_\tau/m_\tau$ is its gamma factor. 
$\frac{d N_{\rm NC}(E'_\nu,E_\nu)}{d E_\nu}\equiv
\frac{1}{\sigma_{\rm NC}(E'_\nu)}  
\frac{d\sigma_{\rm NC}(E'_\nu,E_\nu)}{dE_\nu}$ and 
$\frac{d N_{\rm CC}(E_\nu,E_\tau)}{d E_\tau}\equiv
\frac{1}{\sigma^{\tau}_{\rm CC}(E_\nu)} 
\frac{d\sigma^\tau_{\rm CC}(E_\nu,E_\tau)}{dE_\tau}$ can
be easily computed. The $\tau$ decay distributions 
$\frac{d N_{\rm dec} (E_\tau,E_\nu)}{d E_\nu}$
and $\frac{d \bar N_{\rm dec} (\bar E_\tau,E_\nu)}
{d E_\nu}$  can be found in Refs.~\cite{reno,gaisserbook}.
 
The third term in Eq.~(\ref{eq:nshower}) represents the neutrino
regeneration by NC interactions and the fourth term represents the
contribution from $\nu_\tau$ regeneration, $\nu_\tau
\rightarrow\tau^-\rightarrow\nu_\tau$, describing the energy
degradation in the process.  The secondary $\nu_\mu$ flux from
$\bar\nu_\tau$ regeneration, $\bar\nu_\tau \rightarrow\tau^+
\rightarrow\bar\nu_\tau\, \mu^+\, \nu_\mu$, is described by the last
term where we denote by over-bar the energies and fluxes of the
$\tau^+$.  $\rm Br_\mu=0.18$ is the branching ratio for this decay.
In Eq.~(\ref{eq:tshower}) the first term gives the loss of taus due to
decay and the last term gives the $\tau$ generation due to CC
$\nu_\tau$ interactions. In writing these equations we have neglected
the tau energy loss, which is only relevant at much higher energies.

An equivalent set of equations can be written for the $\bar\nu$
flux density matrix and for the $\tau^+$ flux. Both sets 
are coupled due to the secondary $\nu$ flux term.

We solve this set of ten coupled evolution equations that describe
propagation through the Earth numerically using the matter density
profile of the Preliminary Reference Earth Model and
obtain the neutrino fluxes in the vicinity of the detector
$\frac{d\phi_{\nu_\alpha}(E,\theta)}
{dE\, d\cos\theta}={\rm Tr}[F_\nu(E,L=2R\cos\theta)\, \Pi_\alpha]\,$ .

\begin{figure}[ht]
\begin{center}
\includegraphics[width=3in]{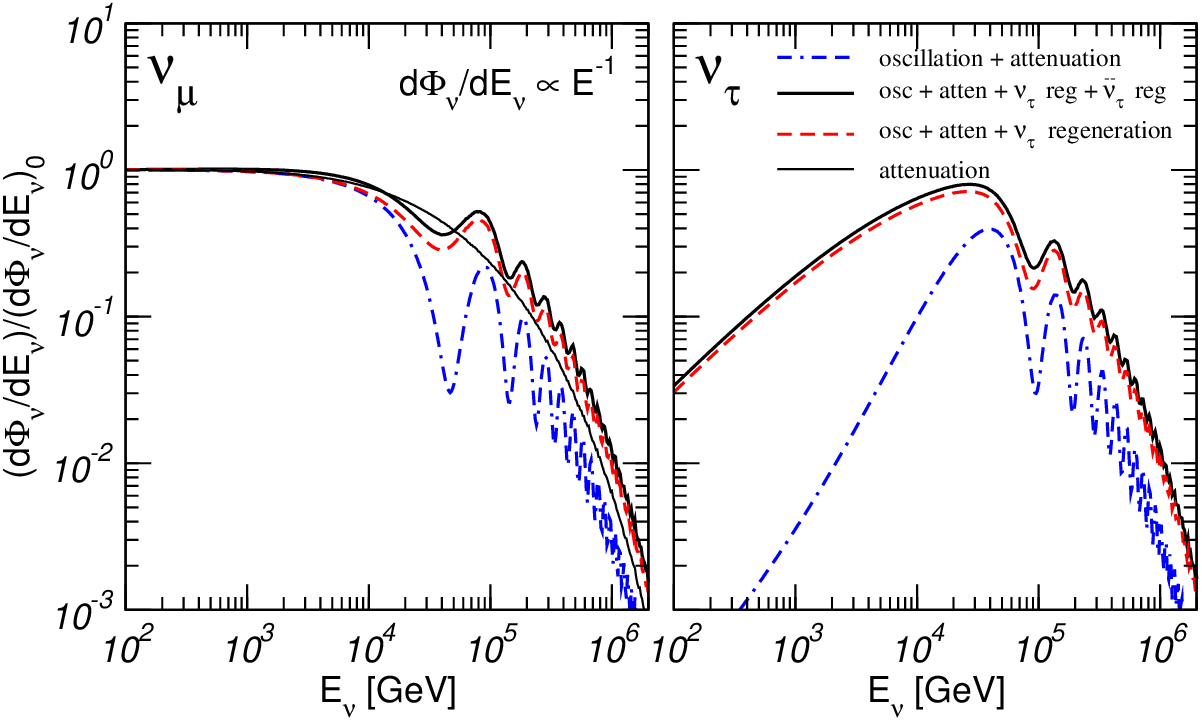}\\
\includegraphics[width=3in]{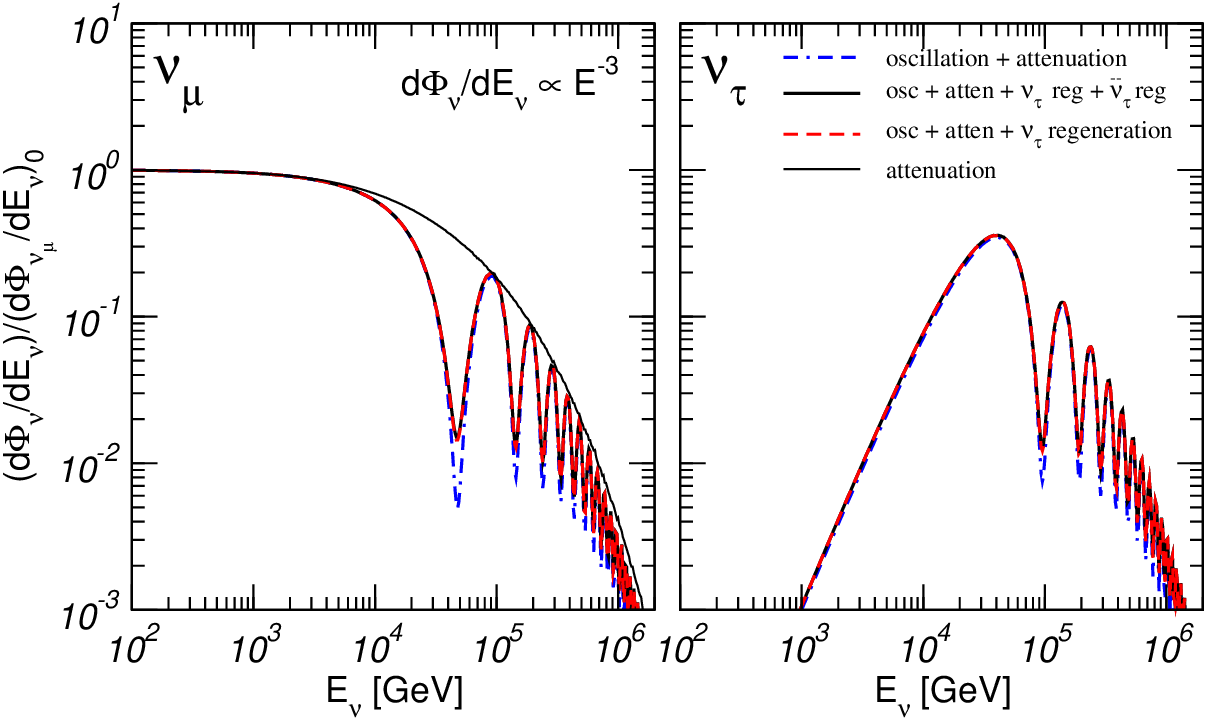}
\end{center}
\caption{\label{fig:neucasc} 
Vertically upgoing neutrinos after traveling the full length of the
Earth taking into account the effects due to VLI oscillations, 
attenuation in the Earth, $\nu_\tau$ regeneration and secondary
$\bar\nu_\tau$ regeneration (see text for details).} 
\end{figure}

In Fig.~\ref{fig:neucasc}  we illustrate the 
interplay between the different terms in Eqs.~(\ref{eq:nshower}) 
and~(\ref{eq:tshower}). The figure covers the example of VLI-induced 
oscillations with $\delta c/c=10^{-27}$ and 
maximal $\xi_{vli}$ mixing. 
The upper panels show the final $\nu_\mu$ and $\nu_\tau$ fluxes
for vertically upgoing neutrinos after traveling the full length of the
Earth for the initial conditions
$d{\Phi(\nu_\mu)_0}/{dE_\nu}=d{\Phi(\bar\nu_\mu)_0}/{dE_\nu}\propto
E^{-1}$ and 
$d{\Phi(\nu_\tau)_0}/{dE_\nu}=d{\Phi(\bar\nu_\tau)_0}/{dE_\nu}=0$.

The figure illustrates that the attenuation in the Earth suppresses
the neutrino fluxes at higher energies. The effect of the attenuation
in the absence of oscillations is given by the dotted thin line 
in the left panel. Even in the presence of oscillations 
this effect can be well described by an overall exponential suppression 
~\cite{gaisserbook,ls} both for $\nu_\mu$'s and the oscillated $\nu_\tau$'s.
In other words, we closely reproduce the curve for 
``oscillation + attenuation" simply by multiplying the initial flux by 
the oscillation probability and an exponential damping factor: 
\begin{equation}
\frac{d\phi_{\nu_\alpha}(L=2R \cos\theta)}{dE d\cos\theta}=
\frac{d\phi_{\nu_\mu,0}}{dE d\cos\theta}\,
P_{\mu\alpha}(E,L=2R \cos\theta) \,
\exp[-X(\theta)(\sigma_{\rm NC}(E)+\sigma_{\rm CC}^\alpha(E))] \, ,
\label{eq:fluxapp}
\end{equation}
where $X(\theta)$ is the column density of the Earth.

The main effect of energy degradation by NC interactions (the third
term in Eq.~(\ref{eq:nshower})) that is not accounted for in the
approximation of Eq.(\ref{eq:fluxapp}) is the increase of the flux in
the oscillation minima (the flux does not vanish in the minimum)
because higher energy neutrinos end up with lower energy as a
consequence of the NC interactions.  The difference between the
dash-dotted line and the dashed line is due to the interplay between
the $\nu_\tau$ regeneration effect (fourth term in
Eq.~(\ref{eq:nshower})) and the flavour oscillations.  As a
consequence of the first effect, we see in the right upper panel that the
$\nu_\tau$ flux is enhanced because of the regeneration of higher
energy $\nu_\tau$'s,
$\nu_\tau(E)\rightarrow\tau^-\rightarrow\nu_\tau(E'<E)$, that
originated from the oscillation of higher energies $\nu_\mu$'s.  In
turn this excess of $\nu_\tau$'s produces an excess of $\nu_\mu$'s
after oscillation which is seen as the difference between the dashed
curve and the dash-dotted curve in the left upper panel.  Finally the
secondary effect of $\bar\nu_\tau$ regeneration (last term in
Eq.~(\ref{eq:nshower})), $\bar\nu_\tau(E)\rightarrow\tau^+\,\rightarrow\mu^+\,\bar\nu_\tau\,\nu_\mu
(E'<E)$, results into the larger $\nu_\mu$ flux  (seen 
in the left upper panel as the difference between the dashed and the thick 
full lines). This, in turn, gives an enhancement in the 
 $\nu_\tau$ flux after oscillations as seen in the right upper panel.

The lower panels show the final 
fluxes for an atmospheric-like energy spectrum
$d{\Phi(\nu_\mu)_0}/{dE_\nu}=d{\Phi(\bar\nu_\mu)_0}/{dE_\nu}\propto
E^{-3}$ and
$d{\Phi(\nu_\tau)_0}/{dE_\nu}=d{\Phi(\bar\nu_\tau)_0}/{dE_\nu}=0$.
In this case  regeneration effects result in the degradation of the neutrino
energy and the more steeply falling the neutrino energy
spectrum, the smaller the contribution to the total
flux. As a result the final fluxes can be
relatively well described by the approximation in
Eq.(\ref{eq:fluxapp}).

\section{Example of Physics Reach: VLI-induced Oscillations}
\label{sec:results}
The expected number 
of $\nu_\mu$ induced events at IceCube can be obtained by 
a semianalytical calculation as:
\begin{eqnarray}
N^{\nu_\mu}_{\rm ev}
&=& T \int^{1}_{-1} d\cos\theta\,  
\int_0^\infty dl'_{min}\, \int^\infty_{l'_{min}} dl\,
\int_{m_\mu}^\infty dE_\mu^{\rm fin}\,
\int_{E_\mu^{\rm fin}}^\infty dE_\mu^0\, 
\int_{E_\mu^0}^\infty dE_\nu \\ \nonumber
&&\frac{d\phi_{\nu_\mu}}{dE_\nu d\cos\theta}(E_\nu,\cos\theta)
\frac{d\sigma^\mu_{CC}}{dE_\mu^0}(E_\nu,E_\mu^0)\, n_T\, 
F(E^0_\mu,E_\mu^{\rm fin},l)\, A^0_{eff}\, \, .
\label{eq:numuevents}
\end{eqnarray}
$\frac{d\phi_{\nu_\mu}}{dE_\nu d\cos\theta}$ is the differential muon
neutrino neutrino flux in the vicinity of the detector after evolution
in the Earth matter obtained as described in the previous section. We
use the neutrino fluxes from Honda~\cite{honda} 
extrapolated to match at higher energies the fluxes from
Volkova~\cite{volkova}.  At high energy prompt neutrinos from charm
decay are important and it is evaluated for   
two different models of charm production: the recombination quark parton
model (RQPM) developed by Bugaev {\sl et al}~\cite{rqpm} and the model
of Thunman {\sl et al} (TIG)~\cite{tig} that predicts a smaller rate.
$\frac{d\sigma^\mu_{CC}}{dE_\mu^0}(E_\nu,E_\mu^0)$ is the differential
CC interaction cross section producing a muon of energy $E_\mu^0$. 
$T$ is the exposure time of the detector. Equivalently,
muon events arise from $\bar\nu_\mu$ 
interactions that are evaluated by an equation similar to
Eq.(\ref{eq:numuevents}).

After production with energy $E_\mu^0$, the muon ranges out in the
rock and in the ice surrounding the detector and looses energy.  We
denote by $F(E^0_\mu,E_\mu^{\rm fin},l)$ the function that describes
the energy spectrum of the muons arriving at the detector.  
We compute the function $F(E^0_\mu,E_\mu^{\rm fin},l)$ by propagating
the muons to the detector taking into account energy losses due to
ionization, bremsstrahlung, $e^+e^-$ pair production and nuclear
interactions according to Ref.~\cite{ls}. 

The details of the detector are encoded in the effective area
$A^0_{eff}$ for which we make a phenomenological parametrization to
simulate the response of the IceCube detector after events that are
not neutrinos have been rejected (
referred to as ``level 2" cuts in Ref.~\cite{IceCube}).  The explicit
form of $A^0_{eff}$ cn be found in Ref.\cite{ouricecube}.

Together with $\nu_\mu$-induced muon events, oscillations also
generate $\mu$ events from the CC interactions of the 
$\nu_\tau$ flux  which reaches the detector producing a
$\tau$ that subsequently decays as 
$\tau\rightarrow \mu \bar \nu_\mu \nu_\tau$ and produces a $\mu$ in the 
detector: 
\begin{eqnarray}
\!\!\!\!\!\!\!\!\!\!\!\!\!\!\!\!
N^{\nu_\tau}_{\rm ev}
&=& T \int^{1}_{-1} d\cos\theta\,  
\int_0^\infty dl'_{min}\, \int^\infty_{l'_{min}} dl\,
\int_{m_\mu}^\infty dE_\mu^{\rm fin}\,
\int_{E_\mu^{\rm fin}}^\infty dE_\mu^0\, 
\int_{E_\mu^0}^\infty dE_\tau 
\int_{E_\tau}^\infty dE_\nu \\ \nonumber
&&
\!\!\!\!\!\!\!\!\frac{d\phi_{\nu_\tau}}{dE_\nu d\cos\theta}(E_\nu,\cos\theta)
\frac{d\sigma^\mu_{CC}}{dE_\tau}(E_\nu,E_\tau)\, n_T\, 
\frac{dN_{dec}}{dE_\mu^0}(E_\tau,E_\mu^0)
F(E^0_\mu,E_\mu^{\rm fin},l)\, A^0_{eff}\,  \; ,
\label{eq:nutauevents}
\end{eqnarray}
where $\frac{d N_{\rm dec} (E_\tau,E_\mu^0)}{d E_\mu^0}$ 
can be found in Ref.~\cite{gaisserbook}. Equivalently we compute
the number of  $\bar\nu_\tau$-induced muon 
events.

Neutrino oscillations introduced by NP effects result in
an energy dependent distortion of the zenith angle distribution 
of ATM muon events. We quantify this effect in IceCube 
by evaluating the expected angular and $E_\mu^{fin}$ distributions in 
the detector using Eqs.~(\ref{eq:numuevents}) and 
~(\ref{eq:nutauevents}) in conjunction with the fluxes obtained after 
evolution in the Earth for different sets of NP oscillation parameters.
\begin{figure}[ht]
\begin{center}
\includegraphics[width=3in]{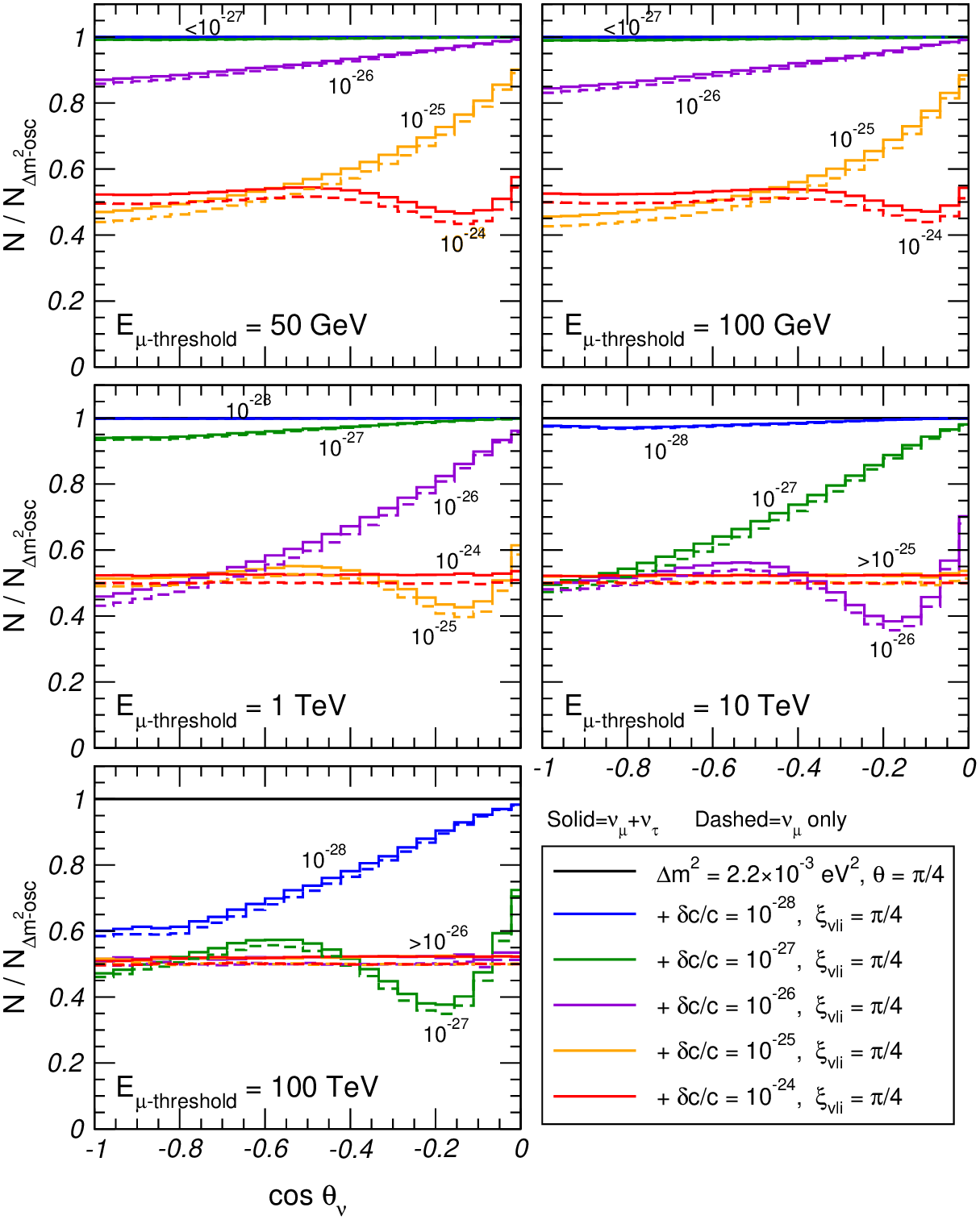}
\includegraphics[width=3in]{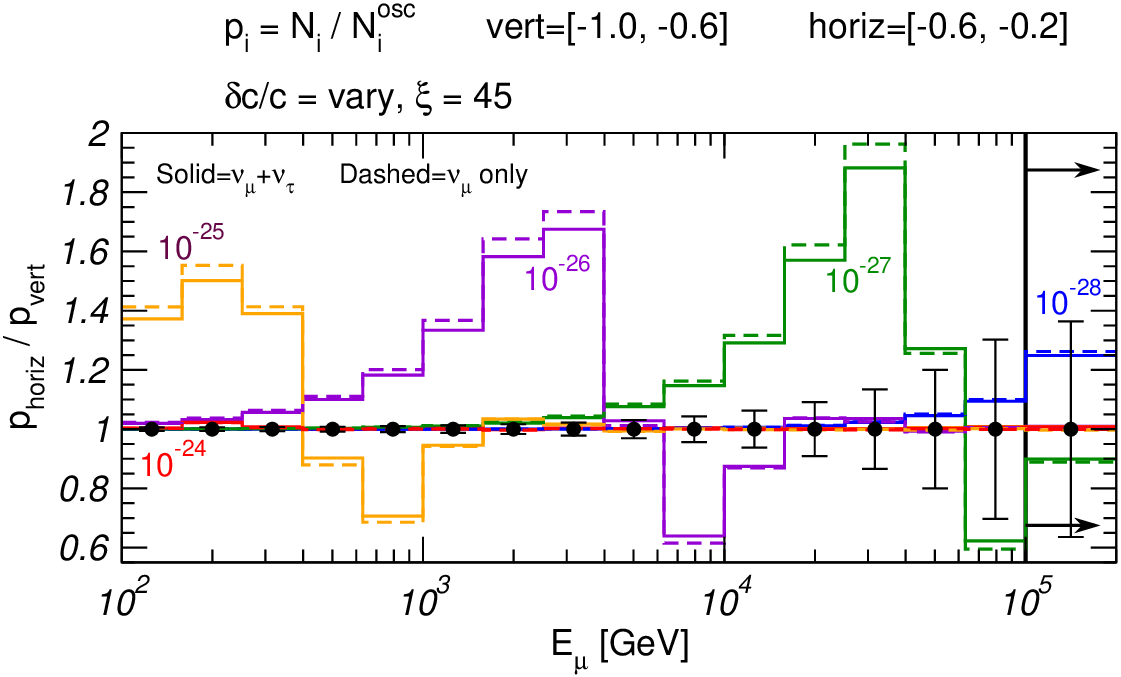}
\end{center}
\caption{\label{fig:zenith} 
{\bf Upper panels}:
Zenith angle distributions for muon induced events for different
values of the VLI parameter $\delta c/c$ and maximal mixing
$\xi_{vli}=\pi/4$ for different threshold energy 
$E_\mu^{fin}>E_{\rm threshold}$ normalized
to the expectations for pure $\Delta m^2$ oscillations . 
{\bf Lower panels}:
The predicted horizontal-to-vertical
double ratio in Eq.(\ref{eq:dblratio}) for different values of $\delta
c/c$. The data points in the figure show the expected statistical
error corresponding to the observation of no NP effects in 10 years of
IceCube.}
\end{figure}

For illustration we concentrate on oscillations resulting from VLI 
that lead to  an oscillation wavelength inversely proportional
to the neutrino energy. The results can be directly applied to 
oscillations due to VEP.   
We show in Fig.~\ref{fig:zenith} the 
zenith angle distributions for muon induced events for different
values of the VLI parameter $\delta c/c$ and maximal mixing
$\xi_{vli}=\pi/4$ for different threshold energy 
$E_\mu^{fin}>E_{\rm threshold}$ normalized
to the expectations for pure $\Delta m^2$ oscillations. 
The full lines include both the $\nu_\mu$-induced events 
(Eq.(\ref{eq:numuevents})) and $\nu_\tau$-induced events
(Eq.(\ref{eq:nutauevents})) while the last ones are not included
in the dashed curves. 
We see that for a given value of $\delta c/c$ there is a range
of energy for which the angular distortion is maximal. Above
that energy, the oscillations average out and result in
a constant suppression of the number of events. 
Inclusion of the $\nu_\tau$-induced events events leads to an overall 
increase of the event rate but slightly reduces the  angular distortion.

In order to quantify the energy-dependent angular distortion we define 
the vertical-to-horizontal double ratio 
\begin{equation}
R_{h/v}(E_\mu^{fin,i})\equiv
\frac{P_{\rm hor}}{P_{\rm ver}}(E_\mu^{fin,i})
=\frac
{\frac
{\displaystyle 
N^{vli}_\mu(E_\mu^{fin,i}, -0.6<\cos\theta<-0.2)}
{\displaystyle 
N^{no-vli}_\mu(E_\mu^{fin,i}, -0.6<\cos\theta<-0.2)}}
{\frac
{\displaystyle 
N^{vli}_\mu(E_\mu^{fin,i}, -1<\cos\theta<-0.6)}
{\displaystyle 
N^{no-vli}_\mu(E_\mu^{fin,i}, -1<\cos\theta<-0.6)}} \; ,
\label{eq:dblratio}
\end{equation} 
where by $E_\mu^{fin,i}$ we denote integration in an energy bin 
of width $0.2\,\log_{10}(E_\mu^{fin,i})$ using that 
IceCube measures energy to 20\% in $\log_{10} E$ for muons. 

In what follows we will use the double ratio in
Eq.~(\ref{eq:dblratio}) as the observable to determine the sensitivity
of IceCube to NP-induced oscillations. We have chosen a double ratio
to eliminate uncertainties associated with the overall normalization
of the ATM fluxes at high energies.  It is worth noticing that
using this observable relies on the fact that the zenith angular
dependence of the effective area is well understood.  

In Fig.~\ref{fig:zenith} we plot the expected value of this ratio
for different values of $\delta c/c$. As mentioned above, IceCube
measures energy to 20\% in $\log_{10} E$ for muons. Accordingly, we
have divided the data in 16 $E_\mu^{fin}$ bins: 15 bins between
$10^2$ and $10^5$ GeV and one containing all events above $10^{5}$
GeV.  In the figure the full lines include both the $\nu_\mu$-induced
events (Eq.(\ref{eq:numuevents})) and $\nu_\tau$-induced events
(Eq.(\ref{eq:nutauevents})) while the last ones are not included in
the dashed curves. As described above, the net result of including the
$\nu_\tau$-induced events is a slight decrease of the maximum expected
value of the double ratio. The data points in the figure show the
expected statistical error corresponding to the observation of no NP
effects in 10 years of IceCube.
In order to estimate the expected sensitivity we assume that no NP effect 
is observed and define a simple $\chi^2$ function including only 
the statistical errors. 

\begin{figure}[ht]
\begin{center}
\includegraphics[width=3.in]{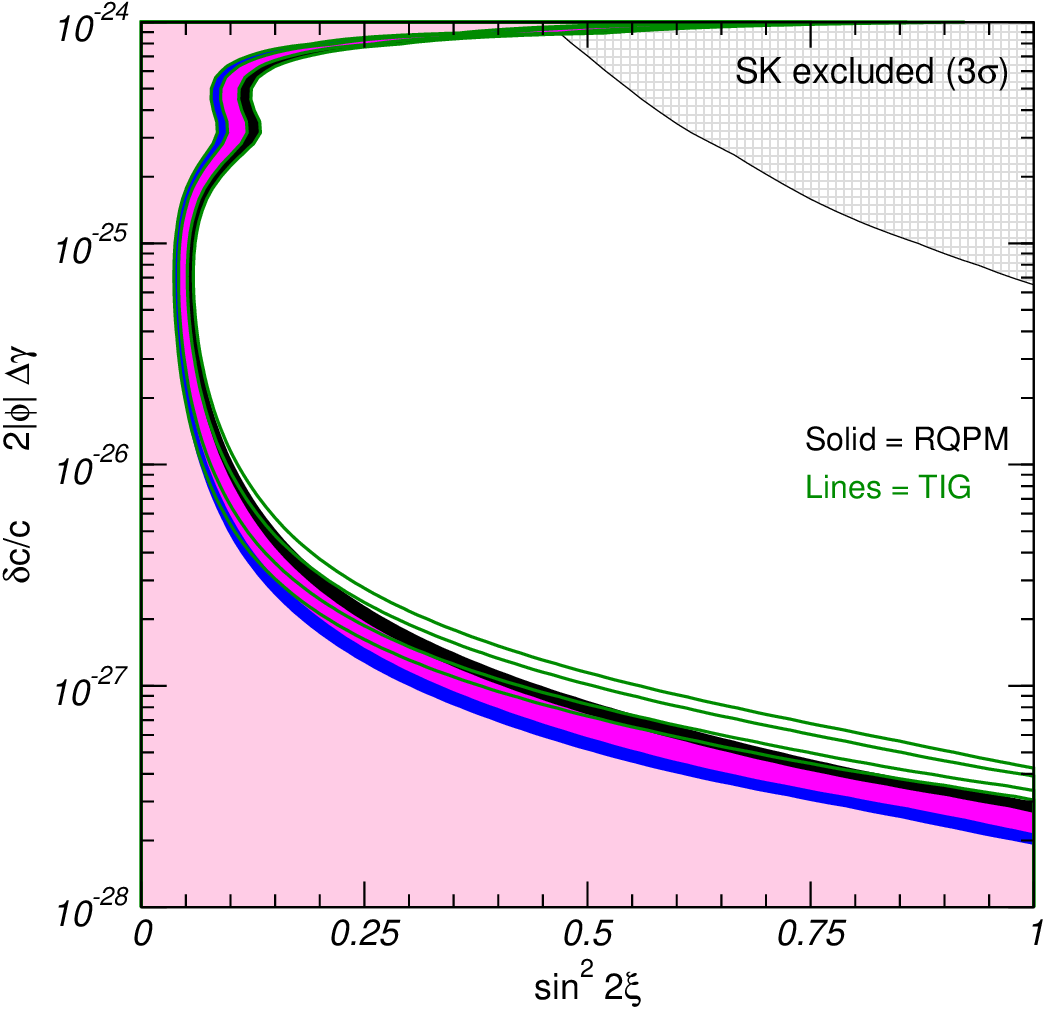}
\end{center}
\caption{\label{fig:chisq} 
Sensitivity limits
in the $\delta c/c, \xi_{\rm vli}$ at 90, 95, 99 and 3 $\sigma$ CL.
The hatched area in the upper right corner is the present $3\sigma$
bound from the analysis of SK data in Ref.~\cite{ouratmnp}.}
\end{figure}

We show in Fig.~\ref{fig:chisq} 
the sensitivity limits
in the $[\delta c/c, \xi_{\rm vli}]$-plane at 90, 95, 99 and 3 $\sigma$ CL
obtained from the condition 
$\chi^2(\delta c/c, \xi_{\rm vli})<\chi^2_{max}({\rm CL,2dof})$.
We show in the figure the results obtained
using the RQPM model and the 
TIG model. The difference is about 
50\%  in the strongest bound on $\delta c/c$. 

The figure illustrates the improvement on the present bounds
by more than two orders of magnitude even within the context of this very 
conservative analysis. The loss of sensitivity at large 
$\delta c/c$ is a consequence of the use of a double ratio as
an observable. Such an observable is insensitive to NP effects 
if $\delta c/c$ is large enough for the oscillations
to be always averaged leading only to an overall suppression.

When data becomes available a more realistic analysis is likely to lead to
a further improvement of the sensitivity. 
\vskip 0.3cm
This work was supported in part by the National Science Foundation
grant PHY0354776 and in part by Spanish Grant No
FPA-2004-00996.

\end{document}